\def\comment#1{}

\newcommand{\curl}{{\rm curl}}
\newcommand{\bfh}{{\bf  H}}

\newcommand{\beg}{\begin{eqnarray}}
\newcommand{\eee}{\end{eqnarray}}

\documentclass[prl,letterpaper,twocolumn,aps,epsf]{revtex4-1}
\usepackage{dcolumn}
\usepackage{amsmath}
\usepackage{mathtools}
\usepackage{amssymb}
\usepackage{graphicx}%
\def\cm#1{}

\begin{document}
\title{
Rotational response of superconductors: \\
magneto-rotational isomorphism and rotation-induced vortex lattice
       }

\author{Egor Babaev${}^{1,2}$ and Boris Svistunov${}^{2,3}$}

\affiliation{
${}^1$ Department of Theoretical Physics, The Royal Institute of Technology 10691 Stockholm, Sweden
\\${}^2$ Department of Physics, University of Massachusetts Amherst, MA 01003 USA
\\ ${}^3$ Russian Research Center ``Kurchatov Institute,'' 123182 Moscow, Russia
}

\begin{abstract}
The analysis of nonclassical rotational response of superfluids and superconductors
was performed  by Onsager (in 1949) \cite{Onsager}  and London (in 1950) \cite{London}  and crucially
advanced by Feynman (in 1955) \cite{Feynman}. It was established that, in thermodynamic limit,
neutral superfluids rotate by forming---without any threshold---a vortex lattice. In contrast, the rotation of superconductors 
at angular frequency ${\bf \Omega}$---supported 
by uniform magnetic field ${\bf B}_L\propto {\bf \Omega}$ due to surface currents---is of the rigid-body type (London Law).
Here we show that, neglecting the centrifugal effects, the behavior of a rotating superconductor
 is identical to that of a superconductor placed in a uniform fictitious external magnetic filed $\tilde{\bf H}=- {\bf B}_L$.
 In particular, the isomorphism immediately implies 
the existence of two critical rotational frequencies in type-2 superconductors.
\end{abstract}

\maketitle
\newcommand{\la}{\label}
\newcommand{\aaa}{\frac{2 e}{\hbar c}}
\newcommand{\Pfaff}{{\rm\, Pfaff}}
\newcommand{\kA}{{\tilde A}}
\newcommand{\G}{{\cal G}}
\newcommand{\cP}{{\cal P}}
\newcommand{\M}{{\cal M}}
\newcommand{\E}{{\cal E}}
\newcommand{\btd}{{\bigtriangledown}}
\newcommand{\W}{{\cal W}}
\newcommand{\X}{{\cal X}}
\renewcommand{\O}{{\cal O}}
\renewcommand{\d}{{\rm\, d}}
\newcommand{\bfi}{{\bf i}}
\newcommand{\e}{{\rm\, e}}
\newcommand{\bfx}{{\bf \vec x}}
\newcommand{\bfn}{{ \vec{\bf  n}}}
\newcommand{\bfs}{{\vec{\bf s}}}
\newcommand{\bfE}{{\bf \vec E}}
\newcommand{\bfB}{{\bf \vec B}}
\newcommand{\bfv}{{\bf \vec v}}
\newcommand{\bfU}{{\bf \vec U}}
\newcommand{\bfp}{{\bf \vec p}}
\newcommand{\f}{\frac}
\newcommand{\bfA}{{\bf \vec A}}
\newcommand{\non}{\nonumber}
\newcommand{\be}{\begin{equation}}
\newcommand{\ee}{\end{equation}}
\newcommand{\ba}{\begin{eqnarray}}
\newcommand{\ea}{\end{eqnarray}}
\newcommand{\bastar}{\begin{eqnarray*}}
\newcommand{\eastar}{\end{eqnarray*}}
\newcommand{\half}{{1 \over 2}}
Superconductors and superfluids  are distinguished by their
Non-Classical Rotational Response (NCRR),  strikingly different  
from the rotational response of ordinary states of matter.
The NCRRs have a high degree of universality rooted in the fact that 
superconductors and superfluid break U(1) symmetry and, at long-length scales,
are described by a complex scalar field theory. { Rotating superfluids form vortex lattices \cite{Onsager,Feynman}, which is routinely used to demonstrate superfluid
properties. 
}

{ By contrast,   a rotational response of a superconductor---known as the London Law \cite{London}---is different.}
It is derived from the minimal, constant-density ($n=const$) model of superconductivity
\ba
F[\phi,{\bf A}]\, &=&\, \frac{n}{4m} \left( \hbar\nabla \phi-\frac{2e}{c}{\bf A} \right)^2 +\frac{1}{8\pi} (\curl {\bf A})^2\,=  \nonumber \\
&& \frac{\gamma n}{2} \left(\nabla \phi+q{\bf A} \right)^2 +\frac{1}{2} (\curl {\bf A})^2,
\label{f0222}
\ea
{ where $F$ is the free energy density, $\phi $ is the superconducting phase, 
 ${\bf A}$ is the vector potential, $m$ and $e$ are the fundamental constants: the mass and charge of the electron.
To simplify notation we employ the units  $\hbar=c=1$, in which $q$ and $\gamma$ are the 
absolute value of bare electric charge and   bare inverse  mass of two electrons.
 According to London \cite{London}, the  superconductor,
rotating with the angular velocity ${\bf \Omega}$,  generates the magnetic field}
\be
 {\bf B}_L\, =\, -\frac{2mc}{e}{\bf \Omega} \, =\,  {2\over \gamma q} \, {\bf \Omega}\,  .
\label{londonlaw}
\ee
{ Current experimental observations  are consistent with the universal character of London's Law (see, e.g., \cite{Hild,London1}). London effect also contributes
to the magnetic field of pulsars, which are rotating protonic superconductors.}
    
Below we show that the state of a rapidly rotating type-2 superconductor becomes different from the London one. The general solution to the problem is readily obtained
by observing that there is an isomorphism between rotating superconductor and a non-rotating superconductor in a uniform external magnetic field.

Assuming electroneutrality  as a natural physical condition and confining ourselves, for simplicity, with the constant-density regime, 
we put superconducting matter field onto a uniformly rotating uniformly charged background. (In superconducting metals, the background charge is associated with the crystal lattice of positively charged ions. In the case of protonic superconductivity in a neutron star, the background charge comes from normal electrons.)
In a neutral superfluid, rotation is equivalent to introduction of a fictitious
vector potential  (and also the centrifugal potential, which we ignore here). Thanks to electroneutrality, the equivalence 
directly applies to our case as well. The key circumstance enforced by electroneutrality  is the invariance of the net electric current 
${\bf J}_{\rm net}={\bf J} + {\bf J}_n$, where ${\bf J}$ is the supercurrent and  ${\bf J}_n$ is electric current of the normal background.
When going to the rotating frame, ${\bf J}_{\rm net}$ remains the same, thus implying the same vector potential field ${\bf A}$.  

In the rotating frame, the free energy density thus reads: 
\be
F[\phi,{\bf A}]\, =\, \frac{\gamma n}{2} \left( \nabla \phi+q{\bf A}- \gamma^{-1}{\bf W} \right)^2 +\frac{1}{2} (\curl {\bf A})^2\, ,
\label{f222}
\ee
with ${\bf W}={\bf \Omega}\times {\bf r}$  the rotating-frame fictitious vector potential.  
{Note that while the uniformly rotating uniformly charged background
does not explicitly enter the free-energy functional, the expression is senseless in the absence of the background.}

For a type-1 superconductor, there will be  a critical rotation frequency
when the London's magnetic field energy density $B_L^2/2$, coming from (\ref{londonlaw}),
becomes equal to superconducting condensation energy. At this rotation frequency, the type-1 superconductor experiences a first-order
phase transition from London regime to a normal state. 
Within Ginzburg-Landau model for a type-1 superconductor, the value of this 
magnetic field coincides with  the  thermodynamical
critical magnetic field  given by $H_{c}=\Phi_0/(4\pi\xi\lambda)$, where $\xi$
and $\lambda$ are coherence and magnetic field penetration lengths, and $\Phi_0$
is the magnetic flux quantum.
Thus 
\be
\Omega_c=\frac{\gamma q\Phi_0}{8\pi\xi\lambda} \, .
\ee
For a type-2 superconductor,  Eq.~(\ref{f222}) allows superconducting state to persist 
at higher rotation frequencies by forming a vortex lattice. The picture becomes immediately clear
by  the following mapping. Introduce {\it shifted} vector potential
\be
\tilde{\bf A} \, =\, {\bf A}- (q\gamma)^{-1}{\bf W}
\ee 
and observe that, in terms of $\tilde{\bf A}$, Eq.~(\ref{f222})  becomes 
isomorphic to free energy of a superconductor in external field $\tilde{\bf H}=-(q\gamma)^{-1}\curl {\bf W}=-{\bf B}_L$:
\be
F[\phi,\tilde{\bf A}] =  \frac{\gamma n}{2} \left( \nabla \phi+q\tilde{\bf A} \right)^2 +\frac{1}{2} \left[ \curl \left(\tilde{\bf A} + {{\bf W} \over q\gamma} \right)\right]^2 \! .
\label{f2222}
\ee
Hence,  type-2 superconductor will have first and second critical rotation frequencies:
\be
\Omega_{c1}=  \f{\gamma q}{8\pi} \left(\f{\Phi_0}{\lambda^2}\right)   \ln \left( \f{\lambda}{\xi}\right) \, ,  \qquad \Omega_{c2}=\frac{\gamma q\Phi_0}{8\pi\xi^2} \, .
\label{omc1}
\ee
{ Restoring dimensional units we get}
\be
\Omega_{c1}=  \f{e}{8\pi m c} \left(\f{\Phi_0}{\lambda^2}\right)   \ln \left( \f{\lambda}{\xi}\right) \, ,  \qquad \Omega_{c2}=\frac{e\Phi_0}{8mc\pi\xi^2} \, .
\label{omc12}
\ee
At $\Omega_{c1}$ the rotating vortex lattice will appear, and at $\Omega_{c2}$ the system will become normal. The London state---
characterized by an ideal diamagnetic response to the
 fictitious magnetic field $\tilde{\bfh}$---emerges as
a counterpart of the Meissner state.   
Under typical conditions, $\Omega_{c1}$ is extremely high. However $\Omega_{c1}(T\to T_c )\to 0$ because $\lambda(T\to T_c )\to \infty$.
{ Also,  $\Omega_{c1}$ can be low enough in thin superconducting films, where the effective penetration length
is  $\lambda_f = 2 \lambda_b^2/d$, with $\lambda_b$  the bulk penetration length and $d$ the film thickness. Given that the length $\lambda_f$ 
can be as large as $\sim$ 1cm, it is clear that attaining the critical value $\Omega_{c1}$  even of order of 10 Hz in thin films is experimentally feasible.
}
A young neutron star is a rotating gradually cooling  system, which undergoes a  superconducting phase transition for protons.
If protonic system forms type-2 superconductor, then  rotation-induced protonic vortex lattices should be generically present in a neutron star at a certain stage of its evolution.

Some physical aspects are worth a discussion.
First, observe that, in the limit  $\Omega \to \Omega_{c2}$, the vortices of the lattice carry  no magnetic flux. 
 In this limit, the fictitious vector potential $(q \gamma)^{-1}{\bf W}$ is 
 compensated by the vortex phase windings in $\nabla \phi$, rather than by ${\bf A}$.
The only relevant vector potential here is ${\bf W}$
and standard Feynman relationship between the flux of corresponding
fictitious field and the vortex density holds.
Physically, the fluxless vortex lattice in this state mimics the solid-body rotation of the superfluid matter field, co-rotating with oppositely charged normal component. 
Apart from possible short-length-scale effects, there is no net transfer of electric charge, ${\bf J}_{\rm net} = 0$,  and thus ${\bf A}=0$.

In a general case, the rotational response will be a combination of 
London response and vortex lattice,  the number of vortices  (antivortices, if negative) inside the system satisfying the relation 
\be
 N = ( \Phi-\tilde{\Phi})/\Phi_0 \, ,
\label{cond1}
\ee
where $\Phi_0=-2\pi/q$ is the magnetic flux quantum,  
 $\Phi$ is the magnetic flux though the system, and  
$\tilde{\Phi}= (q \gamma)^{-1} \oint_{\rm syst.} \! {\bf W} \cdot d{\bf l}~$ is the  flux of the London field ${\bf B}_L$ through the system.
The total magnetic flux per vortex is not quantized due to
the existence of London background field. Nevertheless, an addition of a vortex (antivortex)
{\it at  fixed rotational frequency} amounts to addition  (subtraction) of exactly one flux quantum $\Phi_0$ to (from) 
the total magnetic flux.

The rotational response  of a superconductor is summarized as follows. 
Slow rotation results in creation of uniform magnetic field, in accordance with the London picture. 
The uniformity of the field sets in at the lengths scale $\lambda$ from the boundary of the system.
At the first critical rotation frequency of type-2 superconductor, vortex lattice appears. Each vortex {\it reduces} the total magnetic flux
through the system by one flux quantum---in contrast to the case of Abrikosov lattice \cite{aaa}, where a vortex does the opposite. 
Close to the second critical rotation frequency, the vortex lattice is essentially free of
magnetic flux. 

{
The overall electrical neutrality 
allows one to use similar approach---fictitious gauge field in the rotating frame---in a more general case of multicomponent systems.
For example, in multiband superconductors,  the 
normal background neutralizes
several superconducting components originating in different bands. If, in the rotating frame,
all these components have similar coupling to the fictitious gauge field,  the rotational response of such system
 is isomorphic to its magnetic response, allowing,
in particular,  vortex clusters in the  so-called type-1.5 superconductors \cite{bs1}.
On the other hand, in  the case of a mixture of components with different masses 
and U(1)$\times$U(1) or higher broken symmetry, a rotational response
of the system may be different from the magnetic response because of
different couplings to the fictitious vector potential.}

 We thank Julien Garaud  and Nikolay Prokof'ev for helpful discussions. This work was supported by the Knut and Alice Wallenberg Foundation through a
Royal Swedish Academy of Sciences Fellowship, by the Swedish Research Council, and by the National Science Foundation under the CAREER Award DMR-0955902 and
grant PHY-1314735. We acknowledge the hospitality of the Aspen Center for Physics (supported by the NSF grant 1066293) during the completion of this work.

\end{document}